\documentclass[12pt]{article}
\usepackage{amsmath, mathtools, amsfonts, amssymb, amsthm, bm, float, subcaption}
\usepackage{color}
\usepackage[hidelinks]{hyperref}
\usepackage{setspace}
\usepackage[numbers,sort&compress]{natbib}
\usepackage{graphicx}
\usepackage[letterpaper, left=0.75in, right=0.75in, top=1in, bottom=1in]{geometry}

\title{\LARGE \bf Leading-edge energy transfer governs flutter onset of a pitching finite wing in low-Reynolds number flows}

\author{
    \normalsize Apoorva R. Parvathgari\\
    \small \textit{Department of Mechanical Engineering}\\
    \small \textit{University of Nevada, Reno}\\
    \small Reno, USA\\
    \small aparvathgari@unr.edu
    \and
    \normalsize Tulsi Ram Sahu\\
    \small \textit{Department of Mechanical Engineering}\\
    \small \textit{University of Nevada, Reno}\\
    \small Reno, USA\\
    \small tsahu@unr.edu
    \and
    \normalsize Aditya G. Nair\\
    \small \textit{Department of Mechanical Engineering}\\
    \small \textit{University of Nevada, Reno}\\
    \small Reno, USA\\
    \small adityan@unr.edu
}

\usepackage{titlesec}

\date{}
\begin{document}
\titleformat{\paragraph}
  {\normalfont\normalsize\bfseries}{\theparagraph}{1em}{} 
\titlespacing*{\paragraph}
  {0pt}{3.25ex plus 1ex minus .2ex}{1.5ex plus .2ex} 

\onehalfspacing  

\maketitle

\begin{abstract}
\normalsize
Flutter onset is determined by the sign of the aerodynamic work on the structure, but cycle-integrated work does not reveal where this energy transfer originates. We perform direct numerical simulations of incompressible flow over an aspect-ratio-two NACA 0015 wing pitching at $Re=1000$ and use lobe-resolved force--moment partitioning to identify the vortical contributions to aeroelastic power. The net energy transfer changes sign near $U^*=4.25$. Across this transition, the upper leading-edge contribution changes from damping at $U^*=2$ to excitation near onset and becomes the largest positive contribution at $U^*=6$. Near the stability boundary, positive leading-edge work is nearly balanced by negative trailing-edge work. This change is accompanied by a reorganization of the leading-edge pressure footprint and a reduction in the phase difference between aerodynamic moment and pitch rate, linking flutter onset to the timing of the leading-edge vortex system. A single slowly growing-amplitude simulation recovers the finite-amplitude energy boundary, while a free-pitching calculation confirms the predicted instability. As the motion approaches its limit cycle, the energy pathways redistribute, and a response-frequency-matched forced calculation reproduces the large-amplitude energy budget. These results distinguish the mechanism that triggers flutter from that which saturates the response and identify the leading edge as a natural target for flutter suppression.
\end{abstract}

\section{Introduction}
Aeroelastic flutter results when the fluid performs positive net work on an oscillating
structure. At low Reynolds numbers, this energy exchange is strongly influenced by
separation and vortex shedding, and the resulting instability is not adequately described by a quasi-steady load response. In pitching configurations, the leading-edge flow is particularly important. Vorticity from the leading-edge shear layer can roll up into a coherent leading-edge vortex (LEV), producing a strong suction footprint over the wing. As the LEV grows and convects downstream, the associated pressure loading and centre of pressure change, producing large variations in pitching moment and the moment hysteresis
associated with dynamic stall and stall flutter \citep{2009_dimitriadis_AIAA,2012_mulleners_EF,2013_mulleners_EF,
2020_kissing_EF,2020_li_JFM,2026_martinez_POF}. The LEV growth and trajectory can remain relatively organized during its formation, whereas its saturation and detachment are more sensitive to interactions with secondary vortices and the surrounding separated flow
\citep{2020_kissing_EF,2020_li_JFM,2022_lee_PRF,2023_gardner_PAS,
2026_martinez_POF}. The timing of these events is therefore important not only for the instantaneous aerodynamic loads, but also for whether the resulting pitching moment adds energy to or removes energy from the structural motion.

Energy methods provide a direct measure of this exchange. For prescribed oscillations,
the sign of the cycle-integrated aerodynamic work distinguishes motions that are damped from those that can grow. Such energy maps have been used to predict vortex-induced vibration and stall flutter in two-dimensional configurations
\citep{2009_chkwilliamson_JFM,2010_morse_JFM,2013_shantanu_JFS,2016_samvit_POF,2019_menon_JFM,2020_breuer_JFM}. For a finite wing, however, the same scalar energy measure does not identify the three-dimensional flow structures responsible for the transfer. Leading-edge, trailing-edge and wing-tip vortices interact over the wing and in the near wake, and their topology and coherence can change substantially with the forcing time scale
\citep{2017_visbal_PRF,2020_taira_JFM,2022_lee_PRF,2025_jawahar_JFM}. These changes are reflected in the surface-pressure distribution and hence in the aerodynamic moment. A positive value of the net work therefore establishes instability, but gives little information on where the destabilizing contribution originates.
Moreover, a region producing a large instantaneous moment need not perform positive net work over a cycle. Its contribution to flutter depends on the phase of that moment relative to the pitching motion.

Force--moment partitioning provides a means of making this connection. The method
projects the flow field onto an influence potential and separates physically distinct
contributions to the aerodynamic loads
\citep{2015_rajatmittal_PLOSone,2021_menon_JCP}. It has been used to examine flow-induced vibration and pitching-airfoil aerodynamics \citep{2019_menon_JFM,2021_menon_JFM,2022_menon_JFM,2023_breuer_EF}. For three-dimensional pitching wings, \citet{2023_breuer_JFM} used the partitioned
moment to identify contributions associated with leading-edge and tip-vortex structures.
The relevant quantity for flutter, however, is not the moment alone but the work
performed by each contribution relative to the structural motion. This distinction is
particularly important near the leading edge, where changes in vortex formation and
convection can strongly redistribute the surface pressure and pitching moment.
The remaining question for flutter is to identify which vortical region
changes the aerodynamic energy balance as a finite wing crosses the stability boundary.

We address this question for a NACA~0015 finite wing with $\mathit{AR}=2$ pitching about the mid-chord at $Re=1000$. We resolve the aerodynamic work spatially using force--moment partitioning, with particular attention to the leading- and trailing-edge contributions. The results
show that flutter onset is tied to a change in the leading-edge energy transfer. Near
the stability boundary, positive work from the leading-edge regions is opposed by
trailing-edge damping. On entering the unstable regime, the upper leading-edge lobe
changes sign and becomes the dominant positive vortical contribution. A single
growing-amplitude simulation and a freely pitching calculation are then used to examine how this energy balance evolves beyond onset. The emphasis is on identifying the physical mechanism responsible for the initial loss of stability rather than on a broad
classification of the wake.

\section{Numerical approach and energy partition}

\subsection{Flow configuration}
\begin{figure}
\centering
\includegraphics[width=\textwidth]{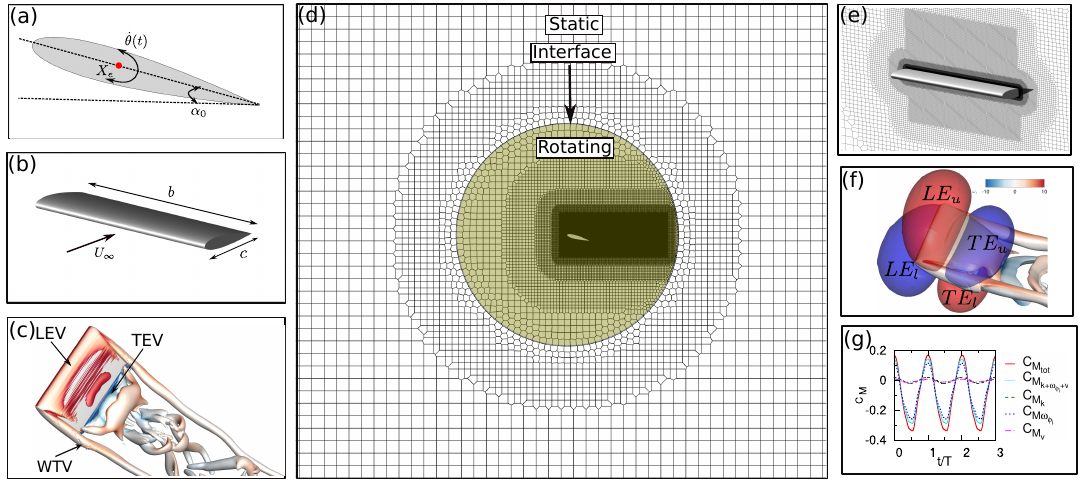}
\caption{Numerical configuration and force--moment partitioning framework.
(\textit{a}) Prescribed pitching about the mid-chord;
(\textit{b}) finite-wing geometry;
(\textit{c}) representative leading-edge (LEV), trailing-edge (TEV)
and wing-tip (WTV) vortical structures;
(\textit{d},\textit{e}) sliding-mesh arrangement and near-wing grid;
(\textit{f}) four lobes of the moment influence potential;
(\textit{g}) representative decomposition of the aerodynamic moment.}
\label{fig:setup}
\end{figure}

We consider an incompressible flow over a aspect ratio $AR=2$, NACA 0015 wing at a mean angle of attack
$\theta_0=15^\circ$, as shown in figure~\ref{fig:setup}(\textit{a},\textit{b}).
The wing pitches about the mid-chord according to
\begin{equation}
\theta(t)=\theta_0+A\sin(2\pi t/U^*),
\label{eq:pitch}
\end{equation}
where $A=5^\circ$ unless otherwise stated and
$U^*=U_\infty/(f_pc)$ is the reduced speed based on the pitching
frequency $f_p$, freestream velocity $U_\infty$ and chord length $c$. Time is normalized by $c/U_\infty$. Simulations are
performed over $0\leq U^*\leq10$ at Reynolds number 
$Re=\rho U_\infty c/\mu=1000$. A representative separated flow field,
showing the leading-edge (LEV), trailing-edge (TEV) and wing-tip (WTV)
vortical structures, is included in figure~\ref{fig:setup}(\textit{c}).

The computations use the finite-volume solver \textit{charLES} with a
rotating near-body mesh coupled to a stationary background grid through
a conservative sliding interface
(figure~\ref{fig:setup}\textit{d},\textit{e}) \citep{2017_bres_AIAA, 2023_bres_AIAA}. The domain extends $20c$ upstream and downstream and $20c$ in the transverse direction, with a spanwise extent of $12c$. The grid contains approximately
$23$ million cells and a near-wing spacing of $2.5\times10^{-4}c$. For the stationary wing at $Re = 400$ and  $\alpha=22^\circ$, the mean lift and drag coefficients are $0.3742$ and $0.3255$, compared with $0.3734$ and $0.3209$ of \citet{2020_taira_JFM}, i.e.\ within $0.21\%$ and $1.43\%$, respectively. At $(Re,U^*,\alpha)=(1000,6,15)$, a grid with approximately twice as many cells changes the mean lift coefficient, $\overline{C_L}$, from $0.3287$ to $0.3373$ ($2.55\%$), the largest difference among all the computed quantities. These checks were used to validate the computational mesh employed in the present study.

\subsection{Energy transfer and lobe-resolved moment}

We use the cycle-averaged fluid-to-structure work rate
\begin{equation}
\overline{E}=\frac{1}{T_f-T_i}
\int_{T_i}^{T_f} C_M(t)\,\dot{\theta}(t)\,\mathrm{d}t,
\label{eq:energy}
\end{equation}
where $C_M$ is the aerodynamic moment coefficient about the pitch axis, and $\dot{\theta}$ is the pitch rate. Here, $T_f-T_i = nT$, where $T$ is the fundamental period of oscillation and $n$ is the number of periods. Positive $\overline{E}$ denotes net energy transfer from the fluid to the structure (negative aerodynamic damping), whereas negative $\overline{E}$ denotes net transfer from the structure to the fluid (positive damping). Statistics for prescribed periodic motions are
evaluated over fully developed cycles after the initial transient.

To localize this energy transfer, we use the force--moment partitioning
method \citep{2015_rajatmittal_PLOSone,2021_menon_JCP}. The moment
influence potential $\phi$ satisfies
\begin{equation}
\nabla^2\phi=0,\qquad
\frac{\partial\phi}{\partial\mathbf n}
=[(\mathbf x-\mathbf x_e)\times\mathbf n]\cdot\mathbf e_z
\end{equation}
on the wing, with a homogeneous Neumann condition at the outer boundary.
The vortex-induced moment over a volume $V$ is proportional to
\begin{equation}
M_V^\omega=-2\int_V Q\phi\,\mathrm{d}V,
\label{eq:fmpm}
\end{equation}
where $Q$ is the second invariant of the velocity-gradient tensor.

For the mid-chord pitch axis, the influence potential separates naturally
into four lobes: the upper and lower leading-edge regions,
$\mathit{LE}_u$ and $\mathit{LE}_l$, and the upper and lower trailing-edge regions,
$\mathit{TE}_u$ and $\mathit{TE}_l$, shown in figure~\ref{fig:setup}(\textit{f}).
We define the corresponding moment and work-rate contributions as
\begin{equation}
C_{M,\phi_i}=\int_{\Omega_i}(-2Q\phi)\,\mathrm{d}V,
\qquad
\mathcal P_{\phi_i}=C_{M,\phi_i}\dot{\theta},
\label{eq:lobe}
\end{equation}
where $\Omega_i$ denotes the $i$th lobe. The cycle average of
$\mathcal P_{\phi_i}$ gives the lobe-resolved energy-transfer rate
$\overline{E}_{\phi_i}$. This decomposition allows the net aerodynamic
work to be traced to specific regions of the separated flow rather than
treated only as a global quantity. A representative reconstruction of the aerodynamic moment from its partitioned contributions is shown in figure~\ref{fig:setup}(\textit{g}). The kinematic ($C_{M_k}$), vorticity-induced ($C_{M_\omega}$) and viscous ($C_{M_v}$) components account for the bulk of the moment, $C_{M_{total}}$, which is the sum of the pressure and viscous contributions.

For the large-amplitude free/forced comparison, we also retain the
complete force--moment partition of the cycle energy into vortex-induced,
pressure, viscous and kinematic contributions. The $-2Q\phi$ contribution
is further separated into $Q>0$ and $Q<0$ regions so that the large
positive and negative terms that cancel at the limit cycle remain visible
rather than being hidden in the net work.

\section{Results}

\subsection{Wake organization and surface loading}

\begin{figure}[tbp]
\centering
\includegraphics[width=\textwidth]{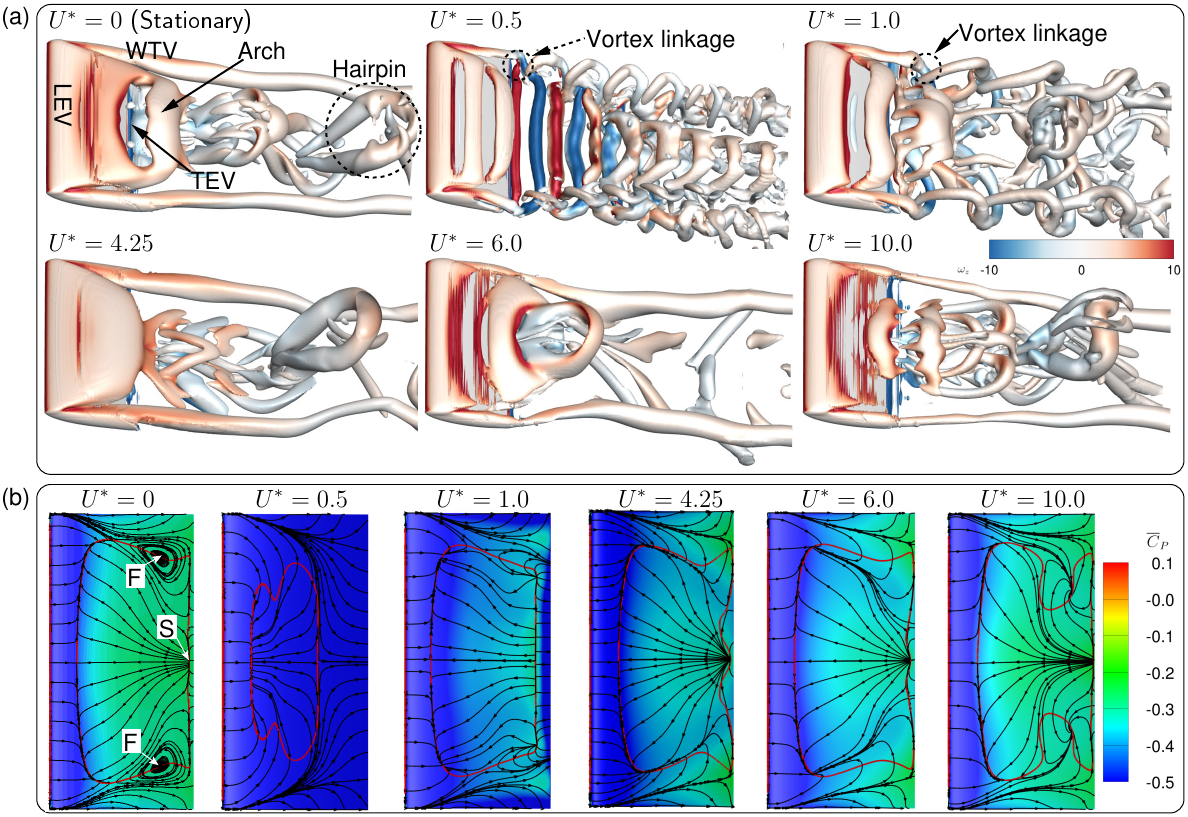}
\caption{Wake organization and surface loading with reduced speed. (\textit{a}) Instantaneous $Q$-criterion isosurfaces for the stationary wing and representative pitching cases, coloured by spanwise vorticity; LEV, TEV and WTV denote the leading-edge, trailing-edge and wing-tip vortices. (\textit{b}) Corresponding time-averaged suction-side pressure coefficient $\overline{C}_p$ with surface streamlines and separation boundary. The wake reorganization is accompanied by a strong redistribution of the leading-edge suction footprint.}
\label{fig:wakepressure}
\end{figure}

The structural response is accompanied by a substantial reorganization of the separated flow over the wing. Figure~\ref{fig:wakepressure}(\textit{a}) shows representative instantaneous vortical structures with increasing reduced speed, $U^*$. For the stationary wing, the separated flow contains a leading-edge vortex together with the wing-tip vortex system and downstream arch- and hairpin-like structures. At low $U^*$, pitching reorganizes the near wake and introduces strong vortex linkage across the span. As the forcing slows with increasing $U^*$, the vortical field undergoes a progressive reorganization and becomes more spatially complex and fragmented. The cases near ($U^*=4.25$) and beyond ($U^*=6$) flutter onset exhibit distinctly different leading-edge and downstream vortex structures.

These changes are reflected in the surface-pressure field in figure~\ref{fig:wakepressure}(\textit{b}). The suction footprint near the leading edge and its downstream extent vary strongly with $U^*$, together with the surface streamline pattern. This connection is important for the pitching dynamics because the aerodynamic moment depends on both the surface loading and its location relative to the pitch axis. A compact change in loading near the leading edge can therefore make a substantial contribution to the moment even when the downstream wake remains energetic. The wake and pressure fields motivate a more direct question: which region of the vortical flow performs positive work on the pitching wing?

The pressure fields alone, however, do not determine whether a particular flow structure stabilizes or destabilizes the wing. This requires the aerodynamic moment to be resolved spatially and evaluated relative to the wing motion. We turn next to the lobe-resolved moment and its contribution to the cycle-averaged work.

\subsection{Leading-edge energy transfer at flutter onset}
\label{sec:onset}
The global energy transfer first establishes the stability change. At $U^*=2$, $\overline{E}=-7\times10^{-4}$ and the prescribed motion is damped. The transfer approaches zero as $U^*$ increases: $\overline{E}=-3\times10^{-4}$ at $U^*=4$ and is approximately neutral at $U^*=4.25$. A case at $U^*=4.5$ gives $\overline{E}=9.62\times10^{-5}$, while $U^*=6$ and $10$ give $1.1\times10^{-3}$ and $6\times10^{-4}$, respectively. The stability boundary therefore lies in the neighbourhood of $U^*=4.25$--$4.5$.

\begin{figure}[tbp]
\centering
\includegraphics[width=\textwidth]{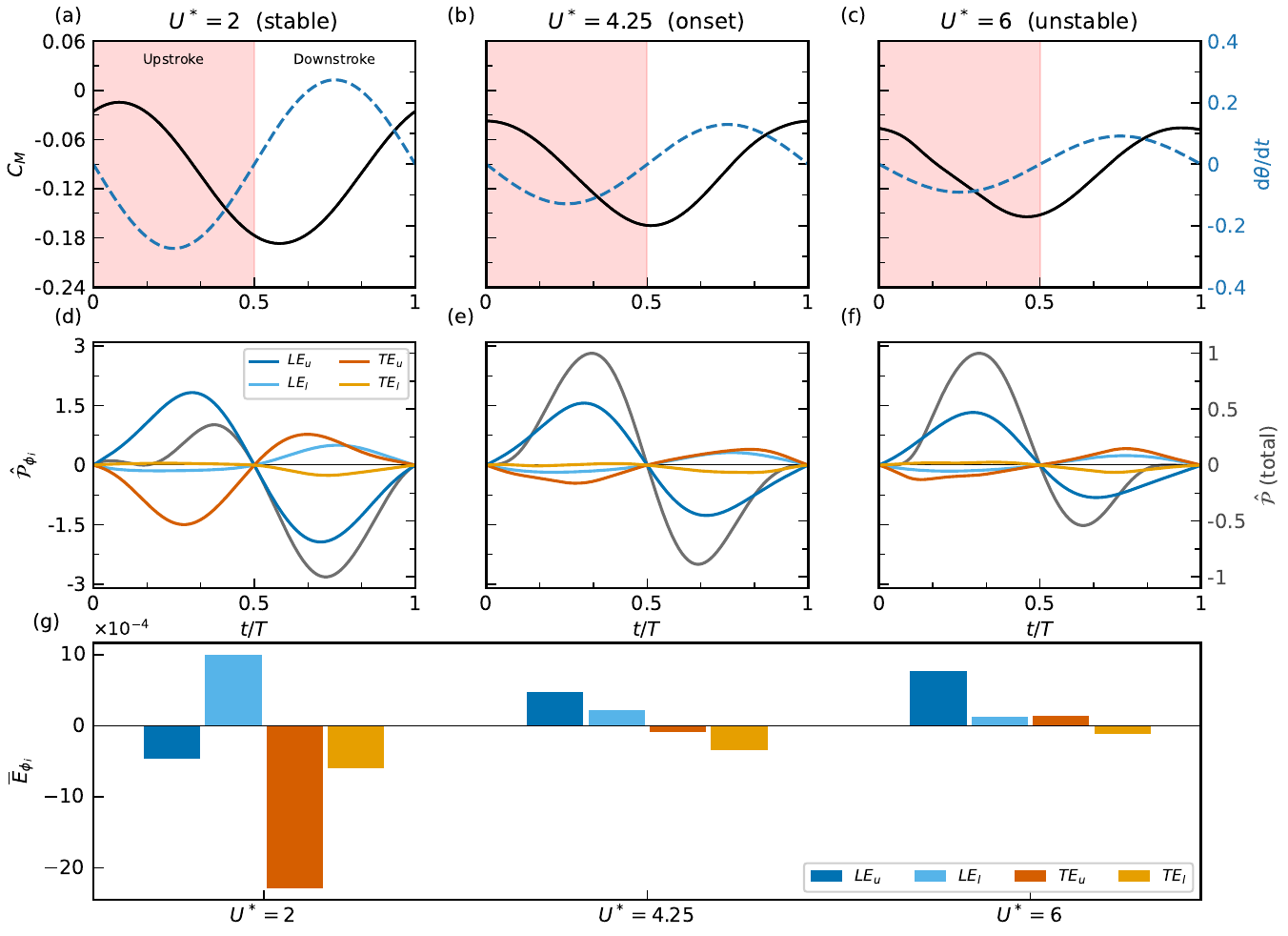}
\caption{Energy transfer across the stability boundary for $U^*=2$ (stable), $U^*=4.25$ (near neutral) and $U^*=6$ (unstable). (\textit{a--c}) Aerodynamic moment $C_M$ and pitch rate $\dot\theta$. (\textit{d--f}) Time-resolved lobe work rates $\hat{\mathcal P}_{\phi_i}$ and total work rate $\hat{\mathcal P}$. Both rates are normalized with their maximum value. (\textit{g}) Cycle-averaged lobe contributions $\overline{E}_{\phi_i}$.}
\label{fig:onset}
\end{figure}

The change in the total work is accompanied by a clear phase change in the aerodynamic moment. Figure~\ref{fig:onset}(\textit{a--c}) compares a stable case ($U^*=2$), a near-neutral case ($U^*=4.25$) and an unstable case ($U^*=6$) for one period of oscillation $T$. The phase difference between $C_M$ and $\dot\theta$ decreases from approximately $94^\circ$ to $90^\circ$ and then to $58^\circ$. The corresponding work rate $\mathcal P=C_M\dot\theta$ consequently changes from net negative to net positive over a cycle. The important point is that the moment amplitude does not vanish at the boundary. Stability is decided by how the unsteady moment is phased with the motion.

The lobe partition identifies the part of the flow responsible for this phase change. The time-resolved lobe contributions in figure~\ref{fig:onset}(\textit{d--f}) show how the work is distributed over each cycle, while figure~\ref{fig:onset}(\textit{g}) summarizes the corresponding cycle-averaged contributions. At $U^*=2$, the upper leading-edge contribution $\overline{E}_{\mathit{LE}_{u}}$ is negative. Its positive and negative portions over the cycle do not cancel, and the net effect is damping. The upper trailing-edge region provides an even larger negative contribution, while $\mathit{LE}_l$ supplies a smaller positive contribution. The stable state is therefore maintained because the negative contributions, particularly from $\mathit{TE}_u$ and $\mathit{LE}_u$, outweigh the positive input from $\mathit{LE}_l$.

Near the boundary at $U^*=4.25$, the balance changes. Both leading-edge lobes contribute positive net work, whereas the trailing-edge contributions are negative. As shown by the cycle-averaged contributions in figure~\ref{fig:onset}(\textit{g}), the total energy remains close to zero because these two regions nearly cancel. The stability boundary is not a state with weak aerodynamic interaction. It is a balance between leading-edge excitation and trailing-edge damping.

At $U^*=6$, the upper leading-edge term increases further and is the largest positive lobe contribution. The lower leading-edge lobe remains positive, $\mathit{TE}_u$ is only weakly positive, and the remaining damping from $\mathit{TE}_l$ is too small to offset the leading-edge input. The leading-edge signature of onset is therefore the sign reversal and growth of the $\mathit{LE}_u$ energy transfer. The simultaneous change in $\mathit{TE}_u$ is secondary in the unstable energy budget. The time-resolved curves in figure~\ref{fig:onset}(\textit{d--f}) show that this is produced by a redistribution of positive and negative work over the cycle rather than by the appearance of a new isolated vortex event. In the flow field, the upper leading-edge lobe is occupied by the leading-edge vortex system. The reduction in phase between $C_M$ and $\dot\theta$ is therefore consistent with a change in the timing of the LEV-induced moment relative to the wing motion. The central mechanism identified here is that the leading edge governs the change from aerodynamic damping to excitation.

\subsection{Amplitude-dependent stability from a single transient}
\label{sec:transient}
The analysis above uses fully developed oscillations at a fixed amplitude. To determine whether the same energy criterion can be obtained without a separate simulation for every amplitude, we prescribe a slowly growing motion at $U^*=6$,
\begin{equation}
\theta(t)=\theta_0+A_0 t\sin(2\pi t/U^*),
\label{eq:transient}
\end{equation}
and evaluate the work over each cycle with $A_0 = 5^\circ$. Two initial conditions are considered: a quasi-steady stationary-wing field (TQS, transient quasi-steady) and a fully developed pitching field at $A=5^\circ$ (TUPW, transient unsteady pitching wing).

\begin{figure}[tbp]
\centering
\includegraphics[width=\textwidth]{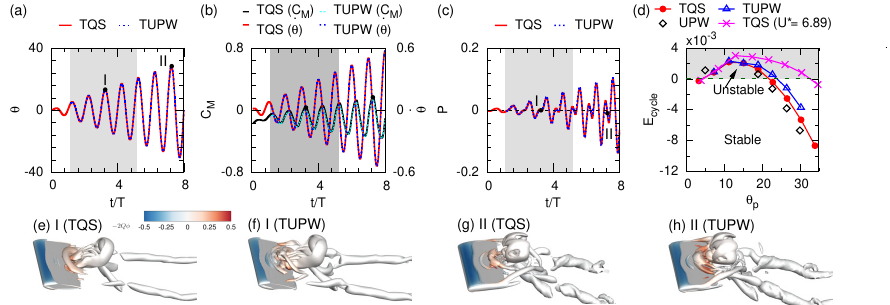}
\caption{Amplitude-dependent energy transfer from a single transient at $U^*=6$. (\textit{a}) Prescribed pitch motion for the two initial conditions TQS (transient quasi-steady) and TUPW (transient unsteady pitching wing); (\textit{b}) aerodynamic moment and pitch rate; (\textit{c}) instantaneous work rate; (\textit{d}) cycle energy versus pitch amplitude compared with fully developed periodic calculations (UPW); (\textit{e--h}) representative $Q$-isosurfaces coloured by the moment density $-2Q\phi$.}
\label{fig:transient}
\end{figure}

The two transients follow nearly the same moment and work-rate histories despite their different initial flow fields (figure~\ref{fig:transient}\textit{a--c}). More importantly, the cycle energy plotted against oscillation amplitude agrees closely with the fully developed fixed-amplitude calculations (figure~\ref{fig:transient}\textit{d}). The energy is positive over an intermediate amplitude range and becomes negative again at large amplitude, giving the finite-amplitude boundary associated with saturation. Thus, at a fixed $U^*$, one transient calculation traces the same stability information that would otherwise require a family of periodic simulations. The instantaneous moment-density fields in figure~\ref{fig:transient}(\textit{e--h}) also show similar leading-edge and near-surface structures for the two initial conditions.

\subsection{Free response and nonlinear saturation}
\label{sec:free}
\begin{figure} [tbp]
\centering
\includegraphics[width=\textwidth]{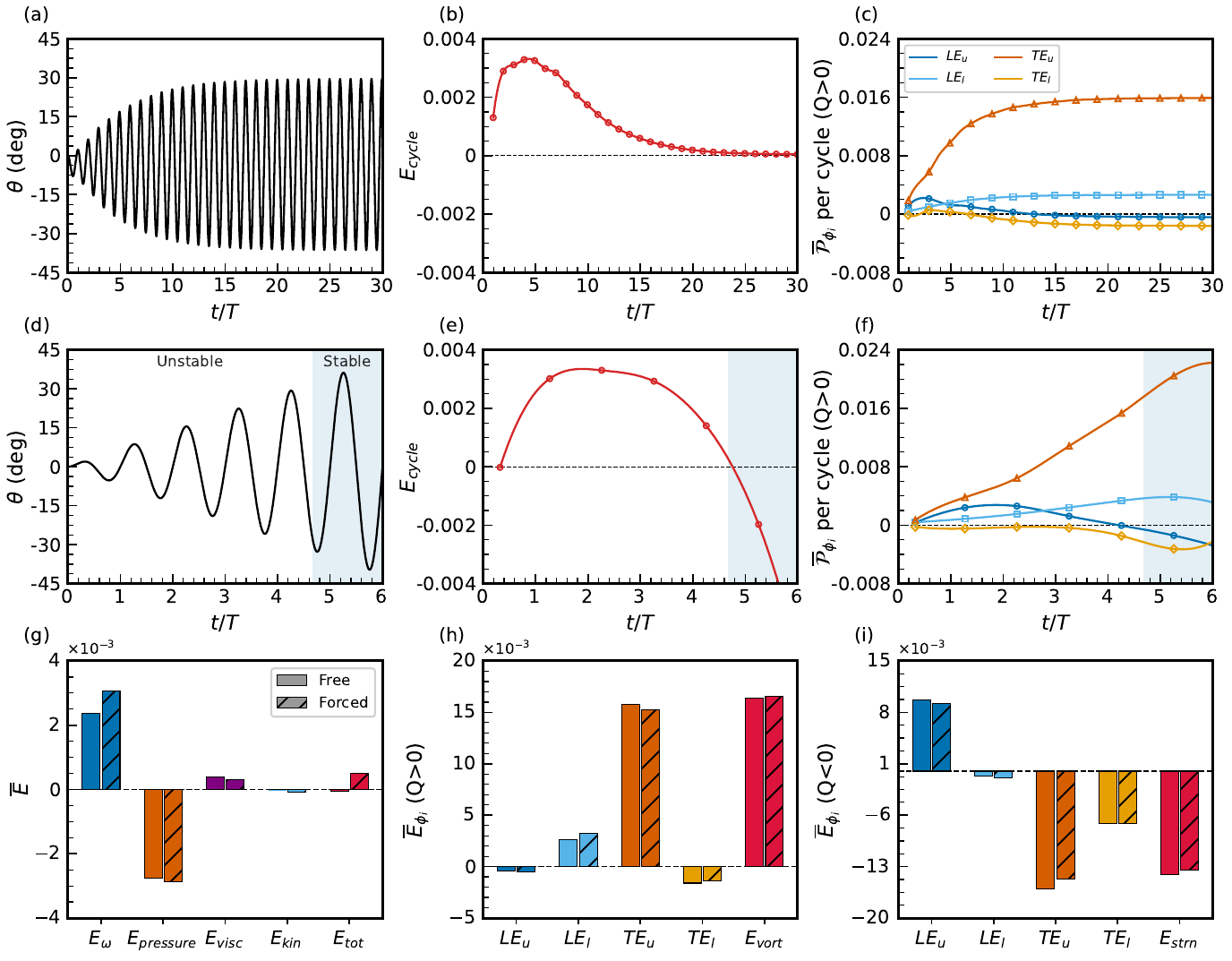}
\caption{Free response, nonlinear saturation and comparison with prescribed motion. (\textit{a}) Growth of the free-pitching displacement at $(Re,U^*,I^*)=(1000,6,1.55)$; (\textit{b}) corresponding cycle energy $E_{\rm cycle}$, which tends to zero as the limit cycle is reached; (\textit{c}) cycle-by-cycle lobe contributions from the $Q>0$ part of the $-2Q\phi$ moment density. (\textit{d--f}) Growing-amplitude prescribed calculation: (\textit{d}) pitch history, with the large-amplitude stable interval shaded; (\textit{e}) cycle energy, which crosses from positive to negative; (\textit{f}) associated $Q>0$ lobe contributions. (\textit{g--i}) Cycle-averaged energy budgets near the saturated response, comparing the free case (solid bars) with the response-frequency-matched forced case (hatched bars): (\textit{g}) vortex-induced, pressure, viscous, kinematic and total contributions; (\textit{h}) lobe-resolved $Q>0$ contributions and their aggregate $E_{\rm vort}$; (\textit{i}) lobe-resolved $Q<0$ contributions and their aggregate $E_{\rm strn}$. }
 \label{fig:free}
\end{figure}

A free-pitching calculation provides a direct check of the prescribed-motion prediction. At $(Re,U^*)=(1000,6)$, the pitch dynamics are
\begin{equation}
I^*\ddot\theta+k^*\theta=C_M,
\end{equation}
with moment of inertia $I^*=1.55$, stiffness $k^*=1.686$ and no structural damping. Starting from the steady state of a stationary wing at $15^\circ$ angle of attack, the oscillation grows and approaches a limit cycle with an amplitude of approximately $28^\circ$ (figure~\ref{fig:free}\textit{a}). The cycle energy is positive during the growth, reaches a maximum during the early transient and then decreases towards zero as the limit cycle is approached (figure~\ref{fig:free}\textit{b}). This is the expected nonlinear closure of the small-amplitude result. At $U^*=6$ the fluid first feeds the motion, but the net work must vanish once a statistically periodic free response is established.

The lobe budget evolves with the amplitude. Figure~\ref{fig:free}(\textit{c}) shows the $Q>0$ contribution from each influence-potential lobe during the free growth. The relative importance of the lobes changes continuously as the oscillation develops, with the positive upper trailing-edge contribution becoming dominant as the limit cycle is approached. The leading-edge result of \S~\ref{sec:onset} should be interpreted specifically as the mechanism that triggers flutter. It does not imply that the same ordering of lobe contributions persists after the wing has reached a large-amplitude periodic state.

Figures~\ref{fig:free}(\textit{d--f}) shows the corresponding growing-amplitude prescribed calculation. The imposed amplitude passes from the positive-energy region into a negative-energy region at large amplitude, and the cycle energy changes sign accordingly. The evolution of the $Q>0$ lobe contributions follows the same broad redistribution seen during the free response. This agreement provides a direct physical check on the transient energy-map construction. The prescribed calculation captures not only the finite-amplitude zero of the net work, but also the way the spatial energy pathways change as that zero is approached.

Because the free oscillator selects a response frequency slightly different from the nominal prescribed frequency, the forced comparison is made at the measured response frequency and at a comparable large amplitude. The comparison is made more stringent in figures~\ref{fig:free}(\textit{g--i}), where the cycle-averaged free response is compared with this response-frequency-matched forced case. The two responses exhibit nearly identical energy budgets near the saturated equilibrium. The complete energy partition in figure~\ref{fig:free}(\textit{g}) shows a large positive vortex-induced contribution opposed primarily by the pressure contribution, leaving only a small net energy at the saturated state. Separating the $-2Q\phi$ term into $Q>0$ and $Q<0$ regions reveals still larger internal cancellations (figures~\ref{fig:free}\textit{h,i}). In particular, the trailing-edge region carries strong contributions of opposite sign in the two partitions. The free and forced budgets remain close in both sign and magnitude across all of these terms. Thus, the onset and saturation are governed by different balances. The leading edge controls the loss of aerodynamic damping at small amplitude, whereas the saturated response is maintained by a redistributed, strongly cancelling three-dimensional energy budget.

\section{Conclusions}
We have examined the fluid-to-structure energy transfer that sets flutter onset for a pitching finite wing at $Re=1000$. The global work changes sign near $U^*=4.25$--$4.5$, but the lobe-resolved force--moment partition shows that the important change occurs at the leading edge. The upper leading-edge contribution changes from damping at $U^*=2$ to excitation near the boundary and becomes the largest positive contribution at $U^*=6$. Near onset, positive leading-edge work is balanced by trailing-edge damping; beyond onset, this balance is lost. The accompanying change in phase between aerodynamic moment and pitch rate indicates that the instability is governed by the timing of the leading-edge vortex system relative to the wing motion. A single growing-amplitude calculation recovers the nonlinear energy boundary. The free response shows how the energy budget reorganizes as the oscillation grows. In particular, the net work tends to zero at the limit cycle while large positive and negative lobe contributions remain. The corresponding forced calculation reproduces this large-amplitude budget when the response frequency is matched. The leading edge therefore controls the onset, whereas saturation is a separate nonlinear balance involving substantial cancellation across the wing. This distinction identifies the leading-edge region as the natural target for localized flutter-control strategies while preserving the distinction between the onset mechanism and the limit-cycle state.

\section*{Acknowledgements}
We thank Karthik Menon and Jawahar Pandi for helpful discussions. Computational resources were provided by the NERSC High Performance Computing Facility, Summit Plus, DoD HPC Modernization program and the Pronghorn High Performance Computing Facility at the University of Nevada, Reno.

\section*{Funding}
AGN acknowledges the support from the Air Force Office of Scientific Research (Award No.~FA9550-23-1-0483; program manager: Dr.~Gregg Abate) and support from the Department of Energy Early Career Research Award (DE- SC0022945).

\bibliographystyle{plainnat}
\bibliography{references}
\end{document}